\documentclass[aps,final,oneside,twocolumn,nobibnotes,nofootinbib,superscriptaddress,showpacs,centertags]{revtex4}
\usepackage{amssymb,amsmath}
\usepackage[cp1251]{inputenc}
\usepackage[T2A]{fontenc}
\usepackage{bm}
\usepackage{array}
\usepackage{graphicx}
\begin{document}

\title{Renormalization-group description of nonequilibrium critical
short-time relaxation processes: a three-loop approximation}

\author{
\firstname{P.~V.}~\surname{Prudnikov},\email{prudnikp@univer.omsk.su}
\firstname{V.~V.}~\surname{Prudnikov},
\firstname{I.~A.}~\surname{Kalashnikov} }
\affiliation{%
Department of Theoretical Physics, Omsk State University,  Pr.Mira 55A, Omsk 644077, Russia
}%

\date{\today}

\begin{abstract}
The influence of nonequilibrium initial values of the order
parameter on its  evolution at a critical point is described using a
renormalization group approach of the field theory. The dynamic
critical exponent $\theta^\prime$ of the short time evolution of a system
with an $n$-component order parameter is calculated within a
dynamical dissipative model using the method of
$\varepsilon$-expansion in a three-loop approximation. Numerical
values of $\theta^\prime$ for three-dimensional systems are determined
using the Pad\'{e}-Borel method for the summation of asymptotic
series.
\end{abstract}


\maketitle

\section{Introduction}

This study is devoted to the influence of nonequilibrium initial
states on the evolution of magnetization $m(t)$ of a ferromagnetic
system at a critical point. As is known \cite{H-H}, anomalous
features in the phenomena of critical dynamics are determined
primarily by the long-range correlation of long-lived fluctuations
of some thermodynamic variables. In this context, subject of
fundamental interest to study is the process of critical relaxation
of a system from an initial nonequilibrium state -- for example,
that created at temperatures much higher than the critical
temperature and, hence, characterized by a short correlation length
-- to a strongly correlated state at the critical point. Janssen et
al. \cite{Janssen} showed that the critical evolution of a system
from the initial nonequilibrium state with a small magnetization
$m_0 = m(0)\ll 1$ displays a universal scaling behavior of $m(t)$
over a short time early stage of this process, which is
characterized by an anomalous increase in magnetization with time
according to a power law. The exponent characterizing this
relaxation process was calculated \cite{H-H} within a
renormalization group approach using the method of
$\varepsilon$-expansion in a two-loop approximation. Later, the
nonequilibrium critical relaxation in a short time regime was
studied within a three-dimensional Ising model using methods of
computer simulation \cite{Zheng}. The results confirmed theoretical
predictions concerning the power character of evolution of the
magnetization of ferromagnetic systems, but the value of an exponent
of $\theta^\prime = 0.108(2)$ determined from these simulations
significantly deviated from the theoretical predictions of $\theta^\prime =
0.130$ (obtained by direct substitution of the parameter
$\varepsilon = 1$ for three-dimensional systems) and $\theta^\prime =
0.138$ (obtained using the Pad\'{e}-Borel method for the summation
of a very short series with respect to $\varepsilon=1$).

\begin{figure}
\includegraphics[width=0.45\textwidth]{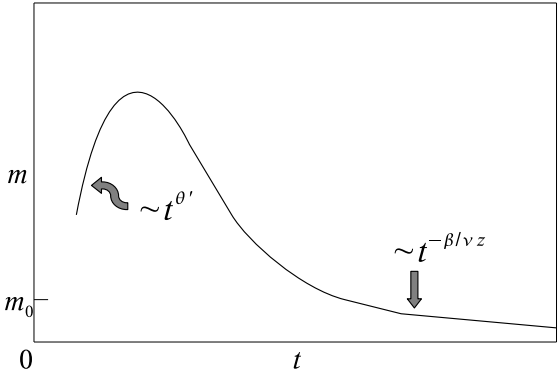}
\caption{ \label{fig:1} Schematic diagram of evolution of
magnetization $m$ at a critical point, starting from an initial
state with magnetization $m_0$.}
\end{figure}

According to the scaling theory, a singular part of the Gibbs
potential $\Phi_{\mathrm{sing}} (t,\tau,h,m_0)$ determining the
state of a system in the critical region is characterized by a
generalized homogeneity with respect to the main thermodynamic
variables:
\begin{equation}
\begin{split}
&\Phi_{\mathrm{sing}}(t,\tau,h,m_0) \\
=b\Phi_{\mathrm{sing}}&(b^{a_t}t, b^{a_\tau}\tau, b^{a_h}h,
b^{a_m}m_0),
\end{split}
\end{equation}
where $t$ is time, $\tau$ is the reduced temperature, $h$ is the
field, $m_0$ is the initial magnetization, $b$ is the scaling
factor, and $a_i$ are the scaling exponents. As a result, the
magnetization $m = - \delta \Phi/\delta h$ of the system at the
critical point ($\tau = 0$, $h = 0$) is characterized by the
following time dependence:
\begin{equation}
\label{m_sceil} m(t, m_0) = t^{-(a_h+1)/a_t}F_m(m_0t^{-a_m/a_t}).
\end{equation}
Expanding the right-hand side of Eq.~(\ref{m_sceil}) into series
with respect to the small parameter $m_0t^{-a_m/a_t}$, we obtain the
following power relation:
\begin{equation}
m(t)\sim t^{-(a_h+a_m+1)/a_t}\sim t^{\theta^\prime}.
\end{equation}
All $a_i$ except $a_m$ can be related to well-known critical
exponents that describe behavior of the system without effects
related to a nonequilibrium initial state. For this reason, Janssen
et al.\cite{Janssen} introduced a new independent dynamic critical
exponent $\theta^\prime$. The renormalization-group description of the
nonequilibrium critical behavior of the system showed that this
exponent takes positive values and, for $t>t_{cr} \sim
m(0)^{-1/(\theta^\prime+\beta/z\nu)}$, the initial regime (characterized by
an increase in the magnetization $m(t)$) changes to a traditional
regime of critical relaxation toward the equilibrium state. The
stage of critical relaxation is characterized by a time dependence
of the magnetization according to the power law $m\sim t^{-\beta/\nu
z}$ (Fig.~\ref{fig:1}), where $\beta$ and $\nu$ are well-known
static exponents determining the equilibrium critical behavior of
the magnetization and the correlation length, and $z$ is a dynamic
exponent characterizing the critical slowing down of relaxation in
the system. It can be shown that, when the system evolves from the
initial ordered state with $m_0 =1$, the time dependence of
magnetization at the critical point is from the very beginning
determined by the power law as $m\sim t^{-\beta/\nu z}$.

\section{Description of the model}

The critical behavior of a pure system in the equilibrium
state is described using the Ginzburg–Landau–Wilson
model Hamiltonian, which can be written as
\begin{equation}
\label{hl}
\begin{split}
H_{GL}[s] = &\int\!\mathrm{d}^{\mathrm{d}}x\,
\Bigg \{ \sum\limits_{\alpha = 1}^n \frac{1}{2!}\left[\left(\nabla s_\alpha(\mathbf{x})\right)^2 + \tau s_\alpha^2(\mathbf{x})  \right] \\
&+\frac{g}{4!}  \left(  \sum\limits_{\alpha = 1}^n s_\alpha^2(\mathbf{x})\right)^2  \Bigg \},
\end{split}
\end{equation}
where
$s(\mathbf{x})$ is the field of the $n$-component order parameter,
$\tau$ is the reduced temperature of the second-order
phase transition, and $g$ is the amplitude of interaction of
order parameter fluctuations.

Let the realization of any configuration of the order
parameter in the system at a given time $t$ be determined
by the condition that the order parameter field $s(\mathbf{x},0)=s_0(\mathbf{x})$
at the initial instant $t=0$ (with the initial magnetization
$m=m_0$) be characterized by the probability distribution
$P[s_0] \sim \exp (-H_0[s_0])$, where
\begin{equation} \label{in}
H_0[s_0] = \int\!\mathrm{d}^{\mathrm{d}}{x}\,\, \frac{\tau_0}{2}\left(s_0(\mathbf{x})-m_0(\mathbf{x})\right)^2.
\end{equation}
In the most interesting case of pure relaxation dynamics
of the order parameter (so-called Model A \cite{H-H}), the
exponent $\theta^\prime$ is essentially new and cannot be expressed
by means of the well-known static critical exponents
and the parameters of equilibrium dynamics. The relaxation
dynamics of the order parameter in this case is
described by the Langevin equation
\begin{equation}
\partial_t s_\alpha(x,t)=-\lambda\frac{\delta H_{GL}[s]}{\delta s_\alpha}+\zeta_\alpha(x,t),
\end{equation}
where $H_{GL}[s]$ is the Ginzburg–Landau–Wilson model
Hamiltonian (\ref{hl}), $\lambda$ is the kinetic coefficient, and $\zeta(x,t)$
is the Gaussian random-noise source, which describes
the influence of short-lived excitations. The randomnoise
source is determined by the following probability
functional:
\begin{gather}
P[\zeta] \sim \exp\left[ -\frac{1}{4\lambda}\int d^dx\int dt(\zeta(x,t))^2 \right]; \nonumber \\[2mm]
\langle\zeta_\alpha(x,t)\rangle = 0;  \\
\langle\zeta_\alpha(x,t)\zeta_\beta(x^\prime,t^\prime)\rangle = 2\lambda\,\delta_{\alpha\beta}\,\delta(x-x^\prime)\delta(t-t^\prime). \nonumber
\end{gather}
Within the framework of the renormalization-group
field theory, the critical dynamics \cite{Lawrie,Bausch} is described in
terms of an auxiliary field $\tilde{s}(\mathbf{x})$ and a generating functional
for the dynamic correlation functions and
response functions. This functional is defined as follows:
\begin{equation}
\begin{split}
W[h,\tilde{h}] &= \ln \Bigg \{ \int \mathcal{D}( s,i \tilde{ s})\exp\left( - \mathcal{L}[ s,\tilde{ s}] - H_0[ s_0]\right) \\
&\times\exp\Biggl (  \int\!\mathrm{d}^{\mathrm{d}}{x}\,\,\int\limits_{0}^{\infty}\!\mathrm{d}{t}\sum_{\alpha=1}^n (\tilde{h}_\alpha\tilde{ s}_\alpha+h_\alpha s_\alpha ) \Biggr ) \Bigg \}, \label{fun}
\end{split}
\end{equation}
where $\mathcal{L}$ is the action functional expressed as
\begin{equation}
\label{1cum}
\begin{split}
\mathcal{L}[ s,\tilde{ s}] &=\int\limits_{0}^{\infty}\!\mathrm{d}{t} \int\!\mathrm{d}^{\mathrm{d}}{x}\,\,\sum_{\alpha=1}^n \Biggl\{\tilde{s}_\alpha\Biggl[\dot{s}_\alpha+\lambda(\tau-\nabla^2)s_\alpha \\
&+\frac{\lambda g}{6}s_\alpha\Bigl( \sum_{\beta=1}^n s_\beta^2\Bigr) -\lambda \tilde{s}_\alpha \Biggr] \Biggr\}.
\end{split}
\end{equation}
An analysis of the Gaussian component of functional
(\ref{1cum}) for $g = 0$ allows the following expressions for the
bare response function $G_0(p,t-t^\prime)$ and the bare correlation
function $C_0^{(D)}(p,t,t^\prime)$ to be obtained for the
Dirichlet boundary condition ($\tau_0 = \infty$) \cite{Janssen}:
\begin{eqnarray}
G_0(p,t-t^\prime)&=&\exp[-\lambda(p^2+\tau)|t-t^\prime|],\\
C_0^{(D)}(p,t,t^\prime)&=&C_0^{(e)}(p,t-t^\prime)+C_0^{(i)}(p,t+t^\prime), \label{cor}
\end{eqnarray}
where
\begin{eqnarray}
C_0^{(e)}(p,t-t^\prime)&=&\frac{1}{p^2+\tau}\,e^{-\lambda(p^2+\tau)|t-t^\prime|}, \label{e} \\
C_0^{(i)}(p,t+t^\prime)&=&-\frac{1}{p^2+\tau}\,e^{-\lambda(p^2+\tau)(t+t^\prime)}.
\end{eqnarray}

\section{Renormalization-group analysis of the model}

In the renormalization-group analysis of the model
with allowance for the interaction of critical fluctuations
in the order parameter, singularities appearing in
the dynamic correlation functions and response functions
in the limit as $\tau \to 0$ were eliminated using the
procedure of dimensional regularization and the
scheme of minimum substraction \cite{Vasiliev} followed by reparametrization
of the Hamiltonian parameters and by
multiplicative field renormalization in functional (\ref{fun}) as
follows:
\begin{equation} \label{ren}
 \begin{array}{ll}
s \to Z_s^{1/2}s,  &\tilde{s} \to Z_{\tilde{s}}^{1/2} \tilde{s}, \\
\lambda \to \left( Z_s/Z_{\tilde{s}} \right)^{1/2}\lambda,  &\tau \to Z_s^{-1}Z_{\tau}\mu^2 \tau,\\
g \to Z_gZ_s^{-2} \mu^\varepsilon g, &\tilde{s}_0 \to \left( Z_{\tilde{s}}Z_0 \right)^{1/2}\tilde{s}_0,
\end{array} 
\end{equation}
where $\varepsilon= 4-d$, and $\mu$ is a dimensional parameter. Calculation
of all renormalization constants $Z_i$ (except
for $Z_0$) was described in \cite{Lawrie}. A scheme for the calculation
of $Z_0$ and the results of calculations in a two-loop
approximation were presented in \cite{Janssen}. In this study, $Z_0$
was calculated in the subsequent three-loop approximation
of the renormalization-group field theory.

\begin{figure*}
\begin{center}
\includegraphics[width=\textwidth]{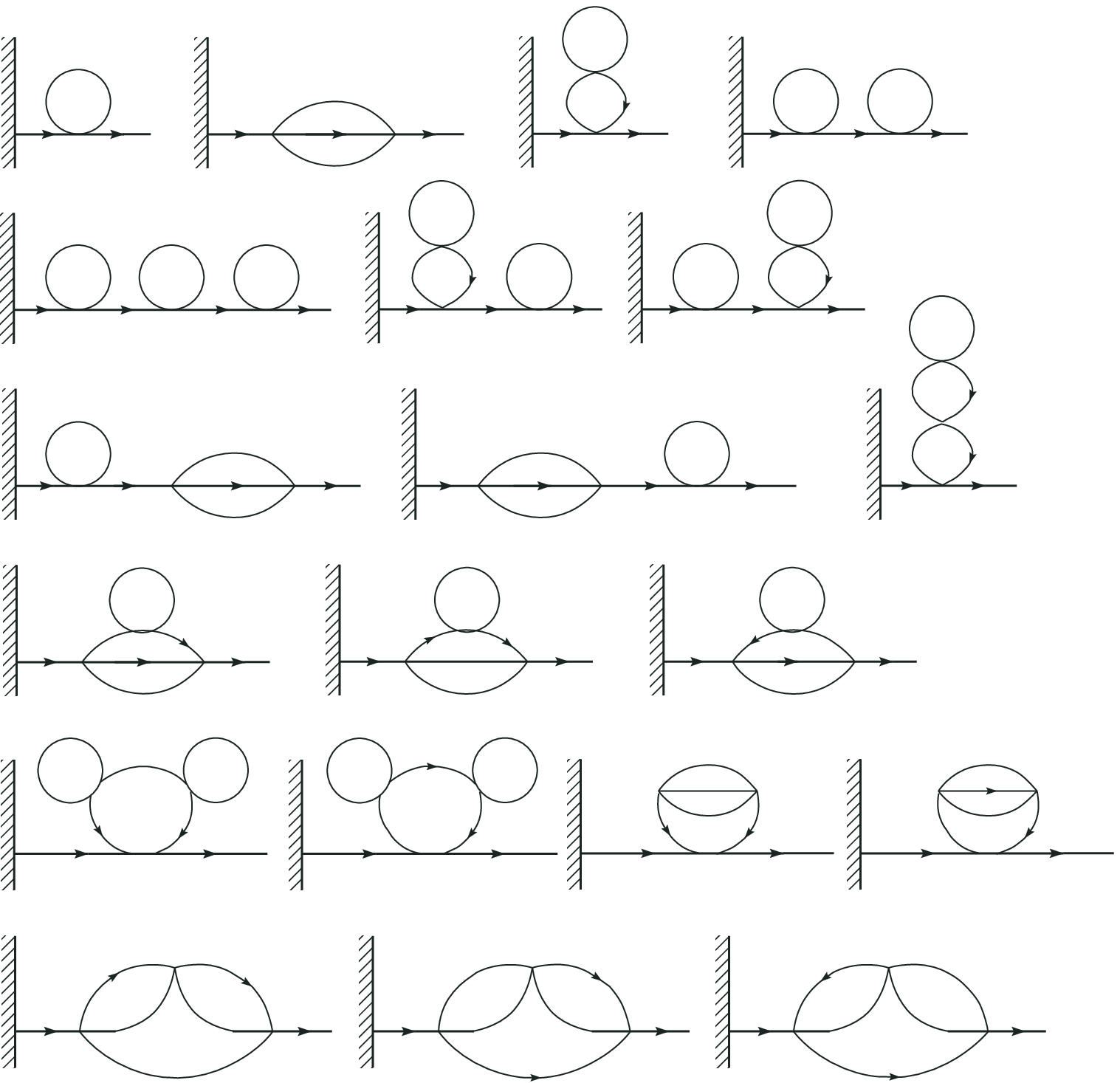}
\caption{ \label{fig:diag} Diagrams determining contributions to the
vertex functions $\Gamma_{1,0}^{(i)}$. Curves without and with
arrows correspond to the bare correlators and the bare response
functions, respectively. The vertical line corresponds to the
surface $t = 0$.}
\end{center}
\begin{center}
\includegraphics[width=\textwidth]{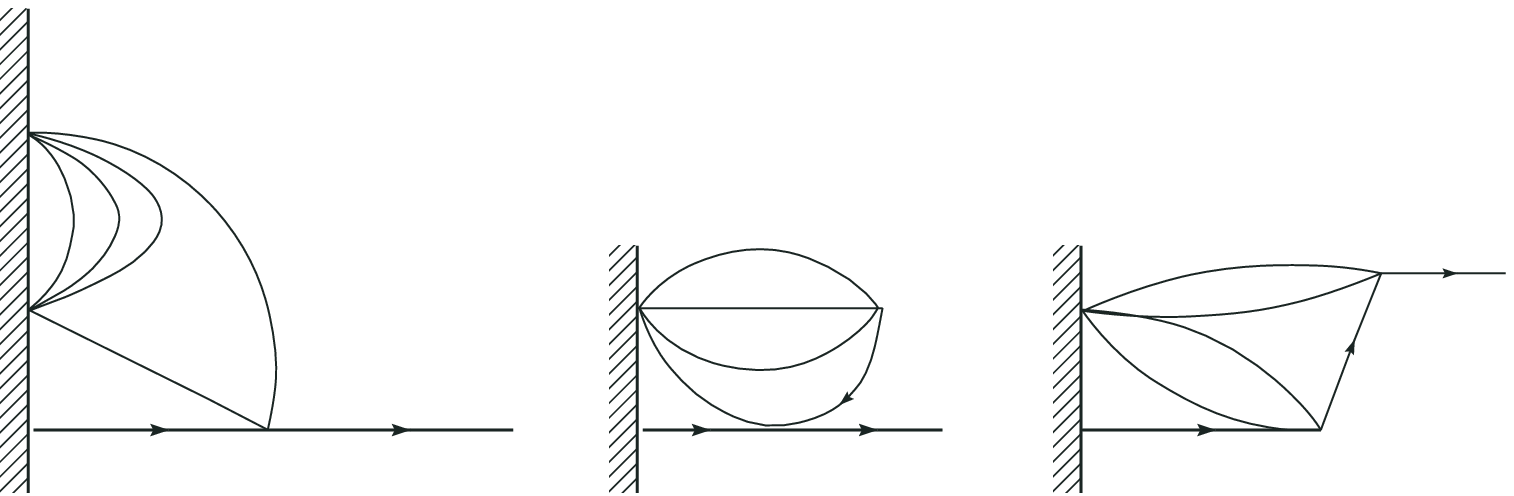}
\caption{ \label{fig:corr} Diagrams determining contributions to the
vertex functions $\Gamma_{1,0}^{(\mathrm{eq})}$.}
\end{center}
\end{figure*}

Introduction of the initial conditions of type (\ref{in}) into the
theory makes it necessary renormalize the response function $\langle
s(p,t)\tilde{s}_0(-p,0)\rangle$, which determines the influence of
the initial state of the system on its relaxation dynamics. The
correction terms in the self-energy part of the response function
appear due to the interaction of fluctuations of the order parameter
and are characterized by reducible dynamic Feynman diagrams, since
they are calculated using correlator (\ref{cor}), which does not
possess the property of translational invariance with respect to
time. Janssen et al. \cite{Janssen} introduced the following
representation for this response function:
\begin{equation} \label{response}
\begin{split}
G_{1,1}^{(i)}(p,t)&=\langle s(p,t)\tilde{s}_0(-p,0)\rangle
\\&=\int\limits_0^t\!\mathrm{d}t^\prime\,\bar{G}_{1,1}(p,t,t^\prime)\,\Gamma_{1,0}^{(i)}(p,t^\prime)_{[\tilde{s}_0]}.
\end{split}
\end{equation}

In a three-loop approximation, the one-particle vertex function
$\Gamma_{1,0}^{(i)}(p,t)_{[\tilde{s}_0]}$ with a single field
insertion $\tilde{s}_0$ is described by the diagrams presented in
Fig.~\ref{fig:diag}, which obey the requirement of containing at
least a single $C_0^{(i)}$ correlator. The factor
$\bar{G}_{1,1}(p,t,t^\prime)$ is determined by the equilibrium
component of the correlator $C_0^{(e)}$ in Eq.~(\ref{cor}). It
should be noted that this correlator differs from the equilibrium
response function $G_{1,1}^{(eq)}(p,t-t^\prime)$ because the
integration in (\ref{response}) with respect to time is performed
starting from $t=0$ instead of $t= -\infty$. However, it is possible
to establish a functional relationship \cite{Oerding} between these
functions by using the functional $H_{GL}[s_0]$ (\ref{hl}) instead
of (\ref{in}) with a new interaction vertex in action functional
(\ref{1cum}):
\begin{equation}
\label{newvert}
\frac{\lambda g}{6} \int\! dt \int\! d^dx\sum_{\alpha,\beta=1}^{n} \left( \tilde{s}_{0\alpha} s_{0\alpha}\right) \left( s_{0\beta}\right)^2.
\end{equation}

The additional vertex function $\Gamma_{1,0}^{(eq)}$, which is
localized on the surface $t=0$, appears due to averaging over the
initial fields. By analogy with representation (\ref{response}), we
also obtain
\begin{equation} \label{Geq}
G_{1,1}^{(eq)}(p,t-t^\prime) = \int_{t^\prime}^{t} dt^{\prime\prime} \bar{G}_{1,1}(p,t,t^{\prime\prime})\Gamma^{(eq)}_{1,0}(p,t^{\prime\prime})_{[\tilde{s}(t^\prime)]}.
\end{equation}
Solving the integral equation
\begin{equation}
 \delta(t-t^\prime) = \int_{t^\prime}^{t} dt^{\prime\prime} K(q,t^{\prime\prime},t^\prime)
\Gamma^{(eq)}_{1,0}(q,t)_{[\tilde{s}(t^{\prime\prime})]}
\end{equation}
in each order of the theory, we obtain the kernel
$K(q,t^{\prime\prime},t^\prime)$. Note that fluctuation corrections
to this kernel in the model under consideration appear only in the
third order (Fig.~\ref{fig:corr}). Using Eqs.~(\ref{response}) and
(\ref{Geq}), performing renormalization of the fields according to
Eq.~(\ref{ren}), and obeying the requirement of eliminating poles
with respect to $\varepsilon$ in each order of the theory so that
the renormalization constant $Z_0$ would remain finite in the limit
as $\varepsilon\to 0$, we eventually obtain the following expression
for this constant:
\begin{equation}
Z_0^{-1/2}\int\limits_0^\infty dt e^{-i\omega t}\hat{\Gamma}_{1,0}(q=0,t)_{[\tilde{s}_0]},
\label{renGam}
\end{equation}
Upon sequential realization of the procedure described above and
calculation of the diagrams using the method of
$\varepsilon$-expansion, the renormalization constant in the
three-loop approximation is as follows:
\begin{multline}
Z_0=1+\frac{n+2}{6}\frac{g}{\varepsilon}+\frac{n+2}{12}\left[\frac{n+5}{3}+\varepsilon\left(\ln 2-\frac{1}{2}\right)\right]\\
\times\left(\frac{g}{\varepsilon} \right)^2+\frac{n+2}{216}\Biggl[ (n+5)(n+6)\\
+\Bigg( \frac{1}{6}(-19+9\ln2)(n+2)+\frac{1}{3}( -7+6\ln2 )
(n+8)\Bigg) \varepsilon{}\\
-1.18679(n+3.13882)\varepsilon^2 \Biggr]\left(\frac{g}{\varepsilon}\right)^3.
\end{multline}

The invariance with respect to the renormalizationgroup
transformations of the generalized connected Green’s function
$$G^{\tilde{M}}_{N,\tilde{N}} \equiv \langle
[s]^N[\tilde{s}]^{\tilde{N}}[\tilde{s}_0]^{\tilde{M}} \rangle$$ can
be expressed in terms of the renormalization-group Callan–Symanzik
differential equation \cite{Janssen,Vasiliev}:
\begin{equation}\label{rg}
\begin{split}
\Bigg\{\mu\partial_\mu+&\zeta\lambda\partial_\lambda+\kappa\tau\partial_\tau+\beta\partial_g+\frac{N}{2}\gamma+\frac{\tilde{N}}{2}\tilde{\gamma}
\\+&\frac{\tilde{M}}{2}(\tilde{\gamma}+\gamma_0)+\zeta\tau_0^{-1}\partial_{\tau_0^{-1}}\Bigg\}G_{N,\tilde{N}}^{\tilde{M}}=0.
\end{split}
\end{equation}
The renormalization-group functions representing the coefficients in
Eq.~(\ref{rg}) are given by the following expressions:
\begin{align}
\label{vf}
\zeta &\equiv(\mu\partial_{\mu})_0\ln\lambda = \dfrac{1}{2}(\tilde{\gamma}-\gamma), &\kappa&\equiv   (\mu\partial_{\mu})_0\ln\tau,\nonumber \\
 \gamma_0 &\equiv (\mu\partial_{\mu})_0\ln Z_0,  &\gamma&\equiv (\mu\partial_{\mu})_0\ln Z_s, \\
\tilde{\gamma} &\equiv  (\mu\partial_{\mu})_0\ln Z_{\tilde{s}}, &\beta &\equiv  (\mu\partial_{\mu})_0 g, \nonumber
\end{align}
where $(\partial_{\mu})_0\equiv \left(\frac{\partial}{\partial
\mu}\right)_0$ denotes differentiation with constant initial
parameters $g$, $\lambda$ and $\tau$. For a short time regime of
nonequilibrium critical relaxation, the only essentially new
quantity is the renormalization-group function $\gamma_0$. In the
three-loop approximation used in this study, this function is
expressed as follows:
\begin{equation}
\begin{split}
\gamma_0=&-\frac{n+2}{6}g \Biggl( 1+\left( \ln 2-\frac{1}{2} \right)g \\
&-0.0988989 \left( n+3.13882\right) g^2 \Biggr)+O(g^4).
\label{gamma0}
\end{split}
\end{equation}
The fixed point $g^*$ of the renormalization-group transformation is
determined from the equation $\beta(g^*)=0$. The general solution of
differential Eq.~(\ref{rg}) by the method of characteristics at the
fixed point has the following scaling form \cite{Janssen}
\begin{align} \label{sc}
&G^{\tilde{M}}_{N,\tilde{N}}(\{x,t\}, \tau,\tau^{-1}_0,\lambda, g^*,
\mu)= l^{(d-2+\eta_s)\frac{N}{2}} \nonumber \\
&\times l^{(d+2+\eta_{\tilde{s}})\frac{\tilde{N}}{2}+(d+2+\eta_{\tilde{s}}+\eta_0)\frac{\tilde{M}}{2}} \\
&\times G^{\tilde{M}}_{N,\tilde{N}}(\{lx,l^{2+\zeta^*}t\},\tau
l^{-2+\kappa^*},\tau^{-1}_0 l^{2+\zeta^*},\lambda, g^*, \mu),
\nonumber
\end{align}
where $\eta_s=\gamma^*$, $\eta_{\tilde{s}}=\tilde{\gamma^*}$ and
$\eta_0=\gamma_0^*$ -- are the exponents of anomalous dimensions.
The functions entering into Eq.~(\ref{sc}) can be related to the
critical exponents involved in the scaling relations, for example:
\begin{align} \label{exp}
z &= 2+\zeta^*, \qquad 1/\nu = 2-\kappa^*, \\
\theta^\prime &= - \left(\zeta^* + \gamma^* + \frac{\gamma_0^*}{2}\right)\left/(2+\zeta^*)\right., \nonumber
\end{align}
which determine the critical relaxation dynamics ($z$), correlation
length ($\nu$), and nonequilibrium evolution ($\theta^\prime$) of the
magnetization. In determining the и value, we used the data of
Kleinert et al.~\cite{Kleinert} on the coordinate of the stable
fixed point $g^*$ and the results of calculations \cite{Antonov} of
the dynamic critical exponent $z$ for the Ising model, which refined
(in the three-loop approximation) the previous values of exponents
for this model. The final expression for the dynamic critical
exponent
\begin{equation}
\begin{split}
z = 2+\frac{\varepsilon^2}{2}\left( 6
\ln\frac{4}{3}-1\right)\frac{n+2}{(n+8)^2}\\
\times\left[1+\varepsilon\left(\frac{6(3n+14)}{(n+8)^2}-0.4384812
\right)  \right],
\end{split}
\end{equation}
represents a generalization of the previous results \cite{Antonov}
to the case of systems with $n$-component order parameters. The
final expression for the critical exponent $\theta^\prime$ is as follows:
\begin{align} \label{th}
\theta^\prime&=\frac{(n+2)}{4(n+8)}\,\varepsilon\Bigg( 1+\frac{6\varepsilon}{(n+8)^2}\left( n+3+(n+8)\ln\frac{3}{2}\right) -\nonumber \\
&-\frac{7.2985}{(n+8)^4}\,\varepsilon^2\Big( n^3+17.3118n^2+153.2670n \\
&+383.5519 \Big)\Bigg)+O(\varepsilon^4). \nonumber
\end{align}

\section{Analysis of results. Conclusions}
The series of $\varepsilon$-expansions exhibit factorial divergence,
but they can be considered in an asymptotic
context \cite{Wilson}. In order to obtain physically reasonable
values of the critical exponents for three-dimensional
systems at $\varepsilon =1$, special methods for the summation
of asymptotic series have been developed \cite{Baker,LeGuillou,Antonenko,Kazakov,Suslov,Prudnikov}, the
most effective being the Pad\'{e}-Borel, Pad\'{e}-Borel-Leroy, and
conformal mapping techniques. We used
the Pad\'{e}-Borel method to perform the asymptotic
summation of $\varepsilon$-expansion series (\ref{th}) for the critical
exponent $\theta^\prime$.
According to this method, the series
\begin{equation}
\theta^\prime(\varepsilon)=\sum\limits_{n=1}^{\infty}c_{n}\varepsilon^{n}
\label{i_0}
\end{equation}
is replaced by the integral
\begin{equation}
\begin{split}
\theta^\prime(\varepsilon)&=\int\limits_{0}^{\infty}e^{-t}B(\varepsilon t)dt, \\ B(x)&=\sum\limits_{n=1}^{\infty}B_{n}x^{n},\qquad B_{n}=\frac{c_{n}}{n!}.
\label{i_f}
\end{split}
\end{equation}
where $B(x)$ is the so-called Borel image. In contrast to
the initial series (\ref{i_0}) having a zero radius of convergence,
the Borel image has a finite radius of convergence
determined by the parameters of the asymptotic
nth term of series (\ref{i_f}) for $n \gg 1$. Then, the Borel image
is subjected to the Pad\'{e} approximation, according to
which $B(x)$ is replaced by a rational function of the following
type:
\begin{equation}
[L/M]=\frac{\sum\limits_{i=0}^{L}a_{i}x^{i}}{\sum\limits_{j=0}^{M}b_{j}x^{j}}\qquad (M\geq 1),
\label{i_[L/M]}
\end{equation}
the expansion of which into Taylor’s series (in the
vicinity of $x = 0$) coincides with that of the Borel image
as far as possible. The function of type (\ref{i_[L/M]}) has $L + 1$
coefficients in the numerator and $M + 1$ coefficients in
the denominator. The entire set of coefficients is determined
to within a constant factor (for certainty, $b_0 = 1$),
so that there are a total of $L + M + 1$ free parameters.
This implies that, in the general case, the coefficients of
expansion of the $[L/M]$ function into Taylor's series
must coincide with the corresponding coefficients of
series (\ref{i_0}). If this series has a finite number $N$ of terms,
the number of coefficients in expansion of the $[L/M]$
function must obey the condition $L+M\leq N$.

In the case of a three-term series under consideration,
the summation of the $\varepsilon$-expansion for exponent $\theta^\prime$
was performed in the $[2/1]$ approximation. In the given
order of the theory, the choice of this approximation is
(in accordance with the results of analysis performed in
\cite{Prudnikov}) preferred for obtaining more accurate value of the
sum. The results of calculations are presented in the
Table~\ref{tab}.

\begin{table}[t!]
\caption{Results of calculations of the critical exponent $\theta^\prime$
for the Ising model and XY model in comparison to the results of
computer simulations} \small \label{tab}
\begin{tabular}{l|cc}\hline\hline
                    &\multicolumn{2}{c}{ Critical exponent $\theta^\prime$}\\
\multicolumn{1}{c|}{Calculation method} & Ising model & XY-model  \\ \hline
\multicolumn{3}{c}{\textbf{ Two-loop approximation}} \\ \hline
 Substitution $\varepsilon = 1$     & $0.130$         & $0.154$ \\
 Pad\'{e}–Borel summation           & \textbf{0.138}  & \textbf{0.170} \\ \hline
\multicolumn{3}{c}{\textbf{ Three-loop approximation}} \\ \hline
 Substitution $\varepsilon = 1$     & $0.0791$        & $0.0983$ \\
 Pad\'{e}–Borel summation           &\textbf{0.1078(22)}  & \textbf{0.1289(23)}\\ \hline
\textbf{Computer simulation}        &\textbf{0.108(2)} \cite{Zheng}&\textbf{0.144(10)}\cite{Kolesnikov}\\
\hline\hline \multicolumn{3}{c}{}
\end{tabular}
\end{table}

A comparison of the results of our calculations of
the critical exponent и to the values obtained by numerical
modeling within the Ising model \cite{Zheng} and XY model
\cite{Kolesnikov} using the method of short time dynamics (see
Table~\ref{tab}) clearly demonstrates that the values obtained in
the three-loop approximation better with the results of
computer simulations as compared to the results of a
two-loop approximation.

In this study, we have presented a field theory
description of the nonequilibrium critical relaxation
of a system within the most interesting dynamical
model A (according to the Hohenberg–Halperin classification \cite{H-H}).
It is shown that only beginning with a
three-loop approximation does the theory of these processes
involve an additional vertex function $\Gamma_{1,0}^{(eq)}$ localized
on the surface of initial states ($t = 0$), which provides
fluctuation corrections to the dynamic response
function due to the influence of nonequilibrium initial
states. As a result, only allowance for these fluctuation
corrections (reflecting the influence of the nonequilibrium
initial states) ensures adequate description of the
relaxation process. Using this three-loop approximation
and the method of $\varepsilon$-expansion, it is possible to
obtain the values of independent dynamic critical exponent
$\theta^\prime$ describing the evolution of the system during a
short time evolution in close agreement with the results
of computer simulations.

\begin{acknowledgments}
This work was supported by the Russian Foundation for Basic Research
through Grants No.~10-02-00507 and No.~10-02-00787 and by Grant No.~MK-3815.2010.2
of Russian Federation President.
\end{acknowledgments}

\end{document}